\documentclass[twocolumn,pra,floatfix]{revtex4}
\usepackage[dvips]{graphicx}
\usepackage{dcolumn}
\usepackage{bm}

\begin{document}

\title{Kinetic energy of uniform Bose-Einstein condensate}
\author{E.~E.~Nikitin$^{1}$ and L.~P.~Pitaevskii$^{2}$}

\begin{abstract}
The Bogoliubov theory of a uniform weakly-nonideal Bose-Einstein condensate
leads to a divergent expression for the kinetic energy of atoms. However,
the latter can be determined provided that the dependence of the scattering
length on atomic mass is known. The explicit expressions are derived for the
kinetic energy through parameters that specify interatomic interaction. The
kinetic energy of condensate atoms noticeably exceeds the total energy.
Numerical data are presented for $^{87}$Rb and $^{23}$Na condensates.
\end{abstract}



\affiliation{$^{1}$~Max-Planck-Institut f\"{u}r
Biophysikalische Chemie,
Am Fassberg, D-37077, Germany and
Department of Chemistry, Technion~-~Israel Institute of Technology,
Haifa, 32000, Israel
\\
$^{2}$Dipartimento di Fisica, Universit\`a di Trento
and BEC Centre, INFM-CNR, Via Sommarive 14, I-38050 Povo,
Trento, Italy;\\ Kapitza Institute for Physical Problems,
ul. Kosygina 2, 119334
Moscow, Russia. }
\maketitle

The Bogoliubov theory of a weakly-nonideal (dilute)
Bose-Einstein condensed gas is known to provide a quite good description of
its properties \cite{Bog}, permitting, in particular, calculation of the
total energy $E$ of the gas in the ground state. However, the Bogoliubov
approach can not be used for the calculation of the total kinetic energy,
the important property of a condensate. Indeed, the distribution function of
atoms out of condensate at large momenta $p$ reads $N_{p}=\left( 4\pi \hbar
^{2}an\right) ^{2}/p^{4}$ where $n$ is the gas density, and $a$ is the
scattering length of slow atoms (see, e.g. \cite{PS}, Sect.~4.3).

Therefore, the total kinetic energy $K$, defined as the integral $K=\int
\frac{p^{2}}{2m}N_{p}\frac{d^{3}p}{\left( 2\pi \hbar \right) ^{3}}$ ,
diverges as $\int dp$ . This means that $K$ is governed by small interatomic
distances, of the order of the interatomic potential range $r_{0}$, where
the Bogoliubov theory is inadequate. The total kinetic energy is an
important quantity that characterizes the interaction, and in principle, it
can be measured (see comments at the end of this letter). The calculation of
this quantity is of considerable theoretical interest along with the total
energy (note, that for an ideal Bose gas at $T=0$, both $E$ and $K$
vanish). We emphasize however that we are speaking here about an uniform
gas in which the kinetic energy is due to the presence of atoms outside the
condensate. This is a non-mean-field effect. An condensate in a trap
possesses also kinetic energy due to its inhomogeneity, which can be
described by the mean-field theory.

In order to calculate the kinetic energy integral, one should use a
microscopic approach that explicitly takes into account properties of
interatomic interaction. This can be done, for $T=0$, by resorting to
well-known formula that relates $K$ to the derivative of $E$ with respect to
the mass of the atom, $m$, viz: $K=-m\left( \partial E/\partial m\right) $
(see \cite{LL5}, Problem to \S~15). For the energy density (the energy per
unit volume) one can use the first-order Bogoliubov expression $E=2\pi \hbar
^{2}n^{2}\left( a/m\right) $ which is valid under condition $na^{3}\ll 1$ ($%
n $ is the density of atoms). In this way we get:
\begin{equation}
K=-2\pi \hbar ^{2}n^{2}m\frac{\partial \left( a/m\right) }{\partial m}\ ,%
\frac{K}{E}=-m\frac{\partial \ln \left( a/m\right) }{\partial m}\ .
\label{General}
\end{equation}
It is thus seen that $K$ can be calculated provided the dependence of $a$ on
$m$ is known. Notice that positiveness of the kinetic energy implies
inequality on this dependence, $\partial \left( a/m\right) /\partial m<0$. A
simple example corresponds to atoms simulated by rigid spheres, when $a$
does not depend on $m$. In this case $K=E$, which is expected for this
example.

We begin from another simple, but instructive model. Let potential of
interaction of atoms can be presented as a spherical square potential well
of radius $R$ and depth $U_{0}$ . Then we have for the scattering length
\begin{equation}
a=R-\frac{\tan \kappa R}{\kappa }\ ,  \label{well}
\end{equation}
where $\kappa =\sqrt{mU_{0}}/m$ (see \cite{LL3}, Problem 1 to \S\ 132).
Simple calculations give
\begin{equation}
\frac{\partial \left( a/m\right) }{\partial m}=-\frac{1}{2m^{2}}\left[
3a+R\kappa ^{2}\left( a-R\right) ^{2}\right] \ .  \label{wellD}
\end{equation}
For a shallow well, when $\kappa R\ll 1$, \ we find
\begin{equation}
\frac{\partial \left( a/m\right) }{\partial m}=-\frac{2U_{0}^{2}R^{5}}{%
15\hbar ^{4}}\ \mathrm{and\ }K=\frac{12\pi \hbar ^{2}n^{2}}{5m}\frac{a^{2}}{R%
}\ .  \label{wellB}
\end{equation}
In this case $K\ll E$~\cite{footnote}. The non-zero value of the kinetic
energy arises only in the second Born approximation. An interesting
situation takes place when $\kappa R\approx \left( N-1/2\right) \pi $, where
$N$ is a positive integer, and when the last term in the r.h.s. of Eq.(2)
makes the main contribution to the scattering length. This is a case, when
the well has a level with small binding energy, $N$ is the level number.
Then $a\gg R$ and one gets from (\ref{wellD}):
\begin{equation}
K=\frac{\pi ^{2}\hbar ^{2}n^{2}}{m}\left( N-1/2\right) ^{2}\frac{a^{2}}{R}\ .
\label{wellF}
\end{equation}

For a more realistic case when the interaction between atoms is described by
a smooth potential $U\left( r\right) $ with a well, the mass dependence of $%
a $ can be calculated within the quasiclassical approximation. The later is
valid, when the number of levels $N$ supported by the well is large and when
the standard quasiclassical approximation for bound states close to the
dissociation threshold is corrected for a ''nonstandard'' matching of the
WKB solution in the well with a quantum solution in the region of the outer
turning point~\cite{Flam}. For our case, the potential in the region of
outer turning is just a van der Waals potential $-\alpha /r^{6}$ where $%
\alpha $ is the van der Waals constant. A relative corrections to this
approximation are of the order of $1/N$, and we consider $1/N$ as a small
parameter of our problem. For alkali dimers, $N$ is of the order of 100.

The quasiclassical expression for the scattering length reads~\cite{Flam}:
\begin{equation}
a=a_{0}\left( 1-\tan \Phi \right)  \label{realG}
\end{equation}
where $a_{0}$ is the ''mean'' scattering length
\begin{equation}
a_{0}=0.48\left( \frac{m\alpha }{\hbar ^{2}}\right) ^{1/4}  \label{a0}
\end{equation}
and $\Phi $ \ is the corrected quasiclassical phase integral
\begin{equation}
\Phi =\frac{1}{\hbar }\int\limits_{r_{i}}^{\infty }\sqrt{-2mU\left( r\right)
}dr-\frac{\pi }{8}\ .  \label{Phi}
\end{equation}
Here $r_{i}$ is the inner turning point that corresponds to the classical
motion with zero energy (the energy here is referred to the dissociation
threshold). Under the quasiclassical conditions, the phase $\Phi $ is large
and is related to $N$ as $\Phi =\pi N+O\left( 1\right) $. Eq.(\ref{realG})
shows a very strong variation of the scattering length upon a change of the
phase integral by the increment of the order of unity. This is expected
since the scattering length is a sensitive function of a position of the
upper, real or virtual, energy level supported by the potential well, see~
\cite{LL3}, \S ~133. In the calculation of the derivative $\partial
a/\partial m$ we can therefore ignore a weak mass dependence of $a_{0}$ as
given by Eq.(\ref{a0}) and take into account the dependence of $\Phi $ from
Eq.(\ref{Phi}) only. After simple derivations we get
\begin{eqnarray}
K &=&\pi ^{2}\hbar ^{2}n^{2}N\frac{a_{0}}{m}\left[ 1+\tan ^{2}\left( \Phi
\right) \right] =  \nonumber \\
&&\pi ^{2}\hbar ^{2}n^{2}N\frac{a_{0}}{m}\left[ 1+\left( a/a_{0}-1\right)
^{2}\right] \   \label{realK}
\end{eqnarray}
Eq.(\ref{realK}) simplifies when the scattering length noticeably exceeds
the ''mean'' scattering length $a_{0}$, that is in the case of the resonance
scattering on a weakly bound state. (Notice, that Bose-Einstein condensate
is stable only if $a>0$ . Hence the level must be a real one.) For $a\gg
a_{0}$ Eq.(\ref{realK}) assumes the form
\begin{equation}
K=\frac{\pi ^{2}\hbar ^{2}n^{2}Na^{2}}{ma_{0}}\ .  \label{KF}
\end{equation}

The appearance of a large factor $N$ in the expression for the kinetic
energy, Eq. (\ref{realK})-(\ref{KF}), reflects the acceleration of relative
motion of a pair of atoms in the potential well. We note that the difference
between $N^{2}$ and $N$ behaviour of $K$ in Eqs. (\ref{wellF}) and (\ref
{realK1}) comes from the fact that a square well and a realistic potentials
show different pattern of convergence of the energy levels to the
dissociation limit. For any realistic potential, the energy levels converge
to the limit in such a way that the spacing between them decreases as $\sqrt{%
-E}$, where $E$ is the energy (negative) counted from the dissociation
limit~ \cite{DN}. The square well does not show this behaviour, since even
for high values of $N$ the bound state is not completely quasiclassical
because of non-quasiclassical behavior of the wave functions close to the
walls.

From Eqs.(\ref{realK}) we get the following expression for the ratio of the
kinetic to total energy:
\begin{equation}
\frac{K}{E}=\frac{\pi }{2}\frac{Na_{0}}{a}\left[ 1+\left( a/a_{0}-1\right)
^{2}\right] \ .  \label{realK1}
\end{equation}
This expression determines the $K/E$ ratio via the scattering length $a$,
''mean'' scattering length $a_{0}$ and the number of bound states $N$. Two
latter quantities, $a_{0}$ and $N$, characterize the microscopic parameters
of the Bose-gas, in addition to the Bogoliubov ''macroscopic'' parameter $a$%
. Rough estimations of $N$ for diatomic molecules can be done using the
model of the Morse potential, $N=2D/\hbar \omega _{e}$ where $D$ is
dissociation energy and $\omega _{e}$ is the small-amplitude frequency of
vibrations of a molecule. Using data from \cite{Smir} we get $N=135$ and $%
N=74$ for $^{87}$Rb$_{2}$ and $^{23}$Na$_{2}$ respectively. Accepting these
values for $N$, calculating $a_{0}$ from (\ref{a0}) with $\alpha $(Rb$_{2}$)$%
=4770$~a.u., $\alpha $(Na$_{2}$)$=1540$~a.u. (see \cite{Smir}), and
borrowing $a$ from~\cite{Rb},~\cite{Na}, we get $a_{0}=79a_{B}$ , $a=109a_{B}
$ for $^{87}$Rb and $a_{0}=43a_{B}$ , $a=52a_{B}$ for $^{23}$Na.
Finally we find $K=176E$ for $^{87}$Rb and $K=100E$ for $^{23}$Na.

It follows from (\ref{realK})-(\ref{realK1}) that the kinetic energy
increases with $a$ faster than the total energy does, and $K$ can be much
more larger then $E$. (The mean-field kinetic energy of inhomogeneous
condensate is always less then the total energy.) In this case, the absolute
value of the total potential energy, $\left| V\right| =\left| E-K\right| $ ,
also exceeds $E$ , so that the total energy emerges from delicate balance of
two large quantities which almost cancel each other. However, it should be
taken into account that an increase in $a$, which happens when the last
bound state approaches dissociation limit, is limited by the condition of a
dilute gas, $na^{3}\ll 1$. Under this condition, the total potential energy
per atom is still much lower than the potential well depth. It is easy to
show for the square well model. Indeed, for an dilute gas, one can conclude
from (\ref{wellF}) that $\left| V\right| /n\sim K/n\sim U_{0}\left(
R/a\right) \ll U_{0}$ . Note that the kinetic energy actually always exceeds
$E$, because the potential energy is negative. Thus the positive total
energy results at the expense of the kinetic energy.

Large values of scattering amplitude have been reached in experiments near
Feshbach resonances \cite{Rob,In}. However in this case one has the
''two-channel situation'', when the resonance level is created by weak
interaction between closed and open channels. This situation demands more
careful theoretical considerations, which will be presented elsewhere.
Notice only that the mass dependence of $a$ for Rb isotopes was investigated
numerically in \cite{Mar}. The kinetic energy of the condensate can be, in
principle, measured. Indeed, according to~\cite{HP} the dynamical formfactor
$S\left( \omega ,q\right) $ of the condensate for a large enough momentum
transfer $q$ yields the momentum distribution; integration of this
distribution allows one to calculate the kinetic energy. It might be more
convenient to use the moments $m_{n}\left( q\right) $ of the formfactor, $%
m_{n}\left( q\right) =\int \omega ^{n}S\left( \omega ,q\right) d\omega $. As
shown in \cite{DS}, one can derive $K$ \ from the asymptotic behavior (for $%
q\rightarrow \infty $) of $m_{2}$ or $m_{3}$.

The dynamical formfactor of the condensate was first measured in \cite{Kett}
by the two-photon Bragg spectroscopy. However, this method does not permit
to get information on the momentum transfer of the order of $\hbar /a_{0}$
which is necessary to recover the kinetic energy. One can get information on
large values of the momentum transfer from neutron scattering data. In this
way the kinetic energy of atoms in liquid $^{4}$He was measured (see, for
example, \cite{He4}). However, this kind of experiments is difficult due to
low density and small dimension of gas. More promising are experiments with
scattering of fast atoms by condensates as discussed in \cite{KS}.
Notwithstanding all difficulties, these types of experiments would open new
important perspectives of investigation BEC at the atomic level.

This theoretical approach can be also compared with results of the Quantum
Monte Carlo calculations. First calculations of the kinetic energy were
presented in \cite{Zhang}. The authors used a model with the delta-potential
interaction and kinetic energy they calculated was related to the gas
inhomogeneity in a trap. However, it is possible to perform such
calculations for a realistic potential.

In conclusion, we have calculated the kinetic energy of a uniform
Bose-Einstein condensate in its ground state. This quantity, which diverges
in the Bogoliubov theory, is determined by the microscopic properties of a
dilute condensed gas: the dependence of the scattering length on the atomic
mass. Different possibility of measurement the kinetic energy of the
condensate are discussed.

We are grateful to S. Stringari who introduced us into the title problem,
C.\ Wieman, S.\ Kokkelmans for constructive comments and B.\ Marcelis for
information about his unpublished results. LPP acknowledges support from
JILA in frame of the JILA Fellowship.


\begin{thebibliography}{99}
\bibitem{footnote}  To ensure condensate stability, one must consider in
this case a potential ''hump'' of height $U_{0}$ instead of the well. Eq. $%
\left( \ref{wellB}\right)$ is valid also for this model.
\end{thebibliography}
\end{document}